# Simulations of Detonation Wave Propagation in Rectangular Ducts Using a Three-Dimensional WENO Scheme

Hua-Shu Dou[1*], Her Mann Tsai[1], Boo Cheong Khoo,[2] Jianxian Qiu[2]
[1]Temasek Laboratories, National University of Singapore, Singapore 117508
[2]Department of Mechanical Engineering, National University of Singapore, Singapore 119260
* corresponding author: FAX: +65 6872 6840; Email: tsldh@nus.edu.sg;
huashudou@yahoo.com

**Abstract** This paper reports high resolution simulations using a fifth-order weighted essentially non-oscillatory (WENO) scheme with a third order TVD Runge-Kutta time stepping method to examine the features of detonation front and physics in square ducts. The simulations suggest that two and three-dimensional detonation wave front formations are greatly enhanced by the presence of transverse waves. The motion of transverse waves generates triple points (zones of high pressure and large velocity coupled together), which cause the detonation front to become locally overdriven and thus form "hot spots". The transversal motion of these hot spots maintains the detonation to continuously occur along the whole front in two and three-dimensions. The present simulations indicate that the influence of the transverse waves on detonation is more profound in three dimensions and the pattern of quasi-steady detonation fronts also depends on the duct size. For a "narrow" duct ($4L \times 4L$ where $L$ is the half reaction length), the detonation front displays a distinctive "spinning" motion about the axial direction with a well-defined period. For a wider duct ($20L \times 20L$), the detonation front exhibits a "rectangular mode" periodically, with the front displaying "convex" and "concave" shapes one following the other and the transverse waves on the four walls being partly out-of-phase with each other.

## I. Introduction

Detonation is a complex phenomenon occurring at supersonic speeds that involve a shock front followed by a reaction zone. Although the study of this phenomenon has a long history, the physics of detonation wave propagation is still an area of active study due to it practical importance [1, 2]. In recent years, the interest in pulse detonation engines has also attracted considerable attention to the study of detonations in tubes or ducts [3, 4]. Both numerical and experimental studies can be found in the literature.

Experimental studies showed that the detonation front has several structures as revealed by the records of soot plate on the sidewalls displaying various cell patterns [4-7]. The structure of detonation front and the feature of cell pattern depend on the duct/pipe size and the properties of the fuels. For the case of detonation occurring in rectangular ducts, studies show the presence of a rectangular mode, diagonal mode and spinning mode [4-7] of the detonation front.

Extensive experiments carried out recently by Hanana et al [8] in rectangular ducts clearly indicate the occurrence of two different modes, namely, rectangular and diagonal detonation front structures. The rectangular structure consists of orthogonal waves traveling independently from each other on the four walls with the triple point lines thus moving parallel to the opposite walls. The soot record shows the classical diamond detonation cell exhibiting "slapping waves." In the case where the transverse waves move along the diagonal line of the duct cross section, diagonal front structures are formed with the triple point formation on a plane normal to the diagonals. The axes of the transverse waves are slanted at $45°$ with respect to the wall, accounting for the lack of "slapping waves" on the wall. Pressure records indicated that where the intensity of the shock front is higher, the averaged wave velocity is larger, and the length of the detonation cell is shorter in forming the diagonal structures. It was suggested that the rectangular mode is effectively a superposition of two-dimensional orthogonal structures. The diagonal structures on the other hand are fundamentally three-dimensional. It was also found that the detonation ignition process is the key parameter controlling the detonation front structure type. However, the relationship between the two different types of structures is unclear, which requires further clarification.

It is known that the experimental diagnostic tools available to study detonation fronts are considerably limited due to the cost the complexity involved. Numerical simulations thus offer advantages in capturing the complex structure of detonation phenomenon and thus act as complements to experiments for the study of detonation propagation. However, the solution of the governing hyperbolic conservation equations that includes chemical reactions, plus the need for high resolution of the flow in numerical modeling makes considerable demands on computing resources. Advancements made in numerical methods coupled with the rapidly decreasing unit cost of computation, makes numerical simulations as a viable tool to explore detonation physics.

Understandably because of high computational expenses, most simulations focus on two-dimensional (2D) detonation [9-14]. However, in recent years three-dimensional (3D) problems [15-19] have been reported. As the physical process taking place in detonation is unsteady and essentially 3D, there are limitations and difficulties in interpretation and understanding detonation physics from purely two-dimensional flow simulations. While there are undoubted similarities of two dimensional simulations with experiments to some extent, the behavior of the flow can best be analyzed and explained with 3D calculations.

Typically many 2D simulations use Eulerian governing equations and one-step Arrhenius chemical reaction models [10-12]. Some calculations are based on multi-step kinetics involving a set of elementary reactions [13-14]. Simulation results generally show that the triple point tracks display a diamond shaped cell pattern. The transverse waves, uncovered by these calculations and found in experiments, seem to play an important role and their trajectory have definite effects on the cell size and also the formation of unreacted fuel pockets [13] due to "engulfment" by the detonation front. It is found that the heat release, activation energy, overdrive factor, and specific heat ratio have significant influence on the detonation stability and the cell pattern shapes [10-12]. The work of Gamezo et al.[11] indicates that for 2D detonation, the average reaction zone is larger and maximum reaction rate is lower than those of the one-dimensional (1D) case. It is suggested that the formation of detonation cells reduces the maximum entropy production in the reaction zone, and hence slows down the system in approaching the equilibrium state. Thus, it is possible that for 3D detonation that such process can be even more pronounced owing to spatial variations of reaction variables. These features can possibly only be analyzed and assessed using 3D simulations.

Williams et al. [15] reported early simulations of 3D detonation in a duct using Euler equations and one-step reaction kinetics. In their simulations, the overdrive factor (the square of the ratio of the detonation velocity of the shock front to the Chapman-Jouguet (CJ) velocity of detonation) was 1.2. Their simulations showed a "rectangular" structure for the detonation front with a phase shift for the front motion between two neighboring walls. It was also shown that the structure of transverse waves is much more complicated than that of the 2D case. Tsuboi et al. [16] also carried out a 3D simulation of detonation in a duct using a more detailed chemical reaction model. They found two types of 3D modes similar to those observed experimentally by Hanana et al [8], namely a rectangular mode and a diagonal mode. An examination of the pattern of maximum pressure history on the wall indicated that the cell length in the rectangular mode is about the same as that for the 2D detonation simulations. On the other hand, the cell length of the diagonal mode is only about three-quarters of that corresponding to the 2D simulations. It was also found that there are more unreacted fuel pockets in the 3D simulations than in the 2D simulations.

Deiterding and Bader [17] also carried out simulations for 3D detonations using a more detailed chemical reaction model. They showed that there is no phase shift between transverse wave directions and that the detonation front featured a single "rectangular mode in phase." All the slices along the transverse direction, the *y*- or *z*-direction in the 3D detonation are similar those of the 2D detonation, or a purely 2D mode with triple point lines just in the y- or z-direction. The period in the 2D and 3D simulations is about the same, but the maximum velocity in the 3D simulation is higher than that in the 2D case.

More recently, He et al. [18] compared 1D, 2D, and 3D detonations using one-step and two-step chemical reaction models and studied the influence of activation energy and energy release on the cell pattern sizes. Their simulations showed that the size of the cell decreases with an increased activation energy but increases with an augmented energy release. For a two-step induction model, the averaged cell length in the 2D case is only about 65% of that from the 3D results.

Thus, there appears to be several inconsistencies among 3D simulation results especially with regard to their relation to the 2D calculations. Further 3D simulations [19] are reported in a rectangular duct using a one-step Arrhenius chemical reaction model and the influence of geometry is considered for three sets of parameters and for periodic and reflection boundary conditions on the walls. The above mentioned two detonation structures were captured by suitably perturbing the initial conditions. It was found that for in-phase rectangular and diagonal structures, there is a similarity in the geometrical evolution of the detonation front. But, the difference of cell length between the two structures seems to be consistent with that found in [16]. In summary most of the above-mentioned 3D simulations focus on describing the structure of the detonation front and few deal with the mechanism of the detonation.

The mechanism of detonation has also been a subject of research interest over the years. It is known that gaseous detonation propagating close to the Chapman-Jouguet (CJ) velocity displays an unstable behavior but the physics of the propagating detonation waves is still not fully understood [3-4]. For example, there is a controversy on the role of transverse waves in the development of the detonation front [4, 9, 20]. One viewpoint is that the transverse waves do not play an essential role in the propagation mechanism. Most of the reactions are induced by the leading shock waves and only a small fraction of the reactions take place directly behind the transverse waves. Another viewpoint is that transverse waves are absolutely essential to the propagation since the collision of transverse waves induces high pressure regions and intensive reaction [20]. It is thus important to clarify the role of transverse waves for detonation particularly in 3D. The recent study by Pintgen et al. [20] seems to support the first viewpoint at least for the gas mixtures considered in that study. However, it does not dismiss the possibility that transverse waves may have an important role in "ordinary" (near CJ) detonations.

The objective of this study is to clarify the mechanism of detonation wave propagation in 3D flow by means of numerical simulations using a high order scheme. This is carried out by comparing the features from 1D, 2D and 3D detonation simulations. In particular, we aim to clarify the role of transverse waves in the detonation sustenance.

Amongst the many methods employed for numerical simulations, the Weighted Essentially Non-Oscillatory (WENO) scheme used recently for detonation studies has allowed the capture of the steep variation of flow properties around the detonation front, as demonstrated in 2D studies performed by Hu et al. [14] and He and Karagozian [21]. We note here that the flow in detonation combustion is supersonic and the propagation of flow is convected by the eigenvectors of the Jacobi matrix of the governing equations. In numerical discretization, it is therefore meaningful that the solution is obtained by solving the flow variables in the direction of eigenvectors of the Jacobi matrix. Although such approach was employed in 2D simulations [14, 21-23], it remains to be demonstrated for 3D problems. As part of our interest to study the detonation mechanism, in this paper we extend the fifth-order WENO scheme to 3D detonation problems to enhance the accuracy of the simulation.

In the following sections, details of the governing equations and numerical method are first described. Then, simulation results are discussed for 1D and 3D detonation with two different duct sizes, and the associated mechanisms of propagation of detonation waves are examined from the simulated flow field.

## 2. Governing Equations and Numerical Method

The governing equations used are described by the three-dimensional Euler equations with a source term that represents chemical reactions. In conservation form, these equations may be written in the compact form

$$\frac{\partial U}{\partial t} + \frac{\partial F}{\partial x} + \frac{\partial G}{\partial y} + \frac{\partial H}{\partial z} = S \tag{1}$$

where the conserved variable vector $U$, the flux vectors $F$, $G$, and $H$ as well as the source vector $S$ are given, respectively, by

$$U = \begin{bmatrix} \rho \\ \rho u \\ \rho v \\ \rho w \\ E \\ \rho Y \end{bmatrix} \quad F = \begin{bmatrix} \rho u \\ \rho u^2 + p \\ \rho uv \\ \rho uw \\ (E+p)u \\ \rho uY \end{bmatrix} \quad G = \begin{bmatrix} \rho v \\ \rho uv \\ \rho v^2 + p \\ \rho vw \\ (E+p)v \\ \rho vY \end{bmatrix} \quad H = \begin{bmatrix} \rho w \\ \rho uw \\ \rho vw \\ \rho w^2 + p \\ (E+p)w \\ \rho wY \end{bmatrix} \quad S = \begin{bmatrix} 0 \\ 0 \\ 0 \\ 0 \\ 0 \\ \omega \end{bmatrix}. \tag{2}$$

Here $u$, $v$, and $w$ are the Cartesian components of the fluid velocity in the $x$, $y$, and $z$ directions, respectively, $\rho$ is the density, p is the pressure, E is the total energy per unit volume, and Y is the reactant mass fraction. The total energy E is defined by

$$E = \frac{p}{\gamma - 1} + \frac{1}{2}\rho(u^2 + v^2 + w^2) + \rho qY, \tag{3}$$

where $q$ is the heat release of reaction, and $\gamma$ is the specific heat ratio. The source term $\omega$ is assumed to be in an Arrhenius form

$$\omega = -K\rho Y e^{-(T_i/T)} \tag{4}$$

where $T$ is the temperature, $T_i$ is the activation temperature, and $K$ is a constant pre-exponential factor. For a perfect gas, the state equation is

$$p = \rho RT. \tag{5}$$

As such, equations (1) to (5) constitute a closed system of equations. The above mentioned equations are made dimensionless based on the state of the unburned gas,

$$\bar{\rho} = \frac{\rho}{\rho_0}, \quad \bar{p} = \frac{p}{p_0}, \quad \bar{T} = \frac{T}{T_0}, \quad \bar{u} = \frac{u}{u_0}, \quad \bar{v} = \frac{v}{u_0}, \quad \bar{w} = \frac{w}{u_0}, \quad \bar{x} = \frac{x}{x_0}, \quad \bar{t} = \frac{t}{t_0}, \quad \bar{E} = \frac{E}{p_0},$$

$$\bar{K} = \frac{Kx_0}{u_0}, \quad \bar{q} = \frac{q}{u_0^2}, \quad \bar{T}_i = \frac{T_i}{T_0},$$

where $u_0 = \sqrt{RT_0}$ and $t_0 = \frac{x_0}{u_0}$. The reference length $x_0$ is chosen as the half-reaction length ($L$), which is defined as the distance between the detonation front and the point where half of the reactant is consumed by chemical reaction. Because of the self-similarity of the Euler equations, the dimensionless form and its original form are identical. For convenience, the overbar on each variable is dropped in the following sections. Here, the one-step reaction model is selected for the study of the essential detonation physics we are concerned with and to avoid complications of multi-step chemical kinetics with possible multi-step reaction models to be explored in the future studies. Moreover, there are published works for this simple model available in the literature for further comparisons.

As the grid used for the present study are regular Cartesian grid, finite volume method or finite difference method can be chosen to solve the governing equations as they both give identical expressions of discretization for the used grid. For brevity, the details of the numerical method used are not included in this paper. It suffice to note that the system of conservation laws of inviscid fluid combined with the one-step chemical reaction model are discretized spatially in eigenvector space using the fifth-order WENO (Weighted Essentially Non-Oscillatory) scheme, and the final discretized variables are solved with a 3rd order TVD Runge-Kutta time integration method [24-25].

The 3D code we have developed (with the chemical reaction term turned off) was first validated for the steady supersonic flow past a wedge and the unsteady flow in one-dimensional shock tubes for the Lax and Sod problems. The cases tested show that the simulated results are in good agreement with those found in the literature [24-25]. The numerical code is further validated using a one-dimensional detonation problem, which will be described below. All these results show that the code is robust and has the desired level of accuracy.

## 3. Results and Discussion
### 3.1. One-dimensional detonation study

Several studies of 1D detonation are reported in the literature [26-31]. In the present work, the 1D detonation simulations serve the following three purposes. Firstly, the 1D simulation is used to validate the code for detonation problems. Secondly, the study can be used to check mesh convergence and determine the required resolution for detonation waves as well as to obtain the CFL (Courant-Friedrichs-Levy) condition applicable to the integration scheme. Thirdly, 1D simulated flow field is used as the initial condition for the simulation of 2D or 3D detonations.

The controlling parameters are the ratio of specific heats γ, the heat release per unit mass of fuel q, the activation temperature $T_i$, and the overdrive parameter $f = (D/D_{cj})^2$. Here, $D$ is the detonation velocity at the front and $D_{cj}$ is the Chapman-Jouguet (CJ) detonation velocity, which can be calculated analytically [32]. One additional free parameter, i.e. the reaction-rate pre-exponential factor $K$, sets the spatial and temporal scales.

A linear stability analysis indicates that the stability of 1D detonation depends on the parameters selected. For detonations at CJ speed, the stability of detonation for a given specific heat ratio depends on the heat release and activation energy. Lee and Stewart [31] indicated that for given f, q, and γ, the detonation becomes more unstable when the activation energy is augmented. For a given q, $T_i$, and γ, the stability of detonation is enhanced when the overdrive factor is augmented. Stability also depends on the magnitude of q. If q is larger than about 5, an increase in q enhances the detonation's stability. If q is less than about 5, increase in q reduces its stability. Eckett et al [33] present a viable criterion for stability on a single curve for a wave traveling at CJ velocity. In the present study, the parameters used for the 1D detonation calculation are $f = 1.0$, q = 50, $T_i$ = 20, and γ = 1.20. For this set of parameters, the simulated neutral curve for instability indicates that the 1D detonation is stable [31, 33].

The 3D code we have developed is used for the 1D detonation simulation studies. The wave propagates from left to right. The boundary condition on the left side of the computational domain is set as reflective and that on the right side is given as a quiescent state. The detonation is initiated by a high pressure on the left of the computational domain. The computational box length in the y and z directions are taken as unity, and the boundary conditions for all variables are zero gradient in the transverse directions. The simulation results are plotted in Fig.1 (a, b) for two different mesh configurations. The corresponding mesh resolution is, 8 and 16 points per half reaction length (*L*) for Fig. 1(a) and (b) respectively. It can be seen that when the time tends to a large value, the detonation becomes stable, which is in agreement with the work performed by Lee and Stewart [31] and Eckett et al. [33]. For both mesh resolution, the peak pressure at the detonation front tends to a constant value of the ZND pressure. Figure 2(a) and (b) display profiles of the final converged solution for various variables, for the two meshes. Owing to the limitation of the mesh size, the detonation velocity obtained is slightly less than the theoretical CJ velocity. The peak pressure is also slightly less than the exact solution of the ZND pressure. We also calculated the 1D detonation using a much finer mesh of 31 points per half reaction length. The results for the mesh convergence for the detonation velocity and peak pressure are shown in Fig. 3(a) and (b), respectively. It can be seen that the detonation velocity converges to the CJ velocity when the mesh size tends to zero, and similarly the peak pressure tends to the theoretical value.

In detonation computations the time step limitation is also determined by the reaction source term besides the Euler convective terms. He and Karagozian [21] used the magnitude of the source term to determine the CFL condition. Here we also adopt the similar approach to evaluate the CFL number for our simulations using a CFL value of 0.2 for 1D simulations and 0.1 for 3D simulations.

**3.2. Three-dimensional detonation study**

The case $f = 1.0$, q = 50, $T_i = 20$, and $\gamma = 1.20$ is first considered as used in several similar numerical studies in one and two dimensions [10, 18, 29, 30]. The initial condition is obtained directly from a precursor 1D simulation (Section 4.1). At time t=0, the flow in both transverse directions is given and set as uniform. The inertial frame of reference moving at the steady CJ detonation velocity is fixed at the shock front. The following boundary conditions are used. The un-burnt fuel mixture enters the domain at an overdrive (supersonic) detonation velocity (here $f$=1). The outflow boundary is set to be non-reflecting; all the parameters at the downstream boundary are extrapolated from the upstream grid points. Since the speed at the outflow boundary is supersonic, the flow behavior downstream does not affect the flow upstream. The walls are modeled by reflective boundary conditions. The present treatments of boundary conditions are consistent and similar to the well-established approach adopted in previous studies [11, 12, 14, 15].

To initiate a disturbance, a random 3D perturbation is added in the form of a localized explosion located immediately behind the leading shock at the first time step. The form of the disturbance is given as $e^* = e + \alpha e g$. Here, $e^*$ is the perturbed total specific energy that encompasses small fluctuations imposed on the reactions, g is a random value ranging from -1 to 1, and $\alpha$ is a coefficient, $0<\alpha<1.0$, that controls the amplitude of the fluctuations [14].

Calculations are carried out for two ducts with different widths. For the narrow duct, the computational box size normalized by the reference length $L$ is $8 \times 4 \times 4$. The number of grid points employed in the simulation was decided by initial test runs and the results reported here uses 121x61x61, which corresponds to a resolution of 15 points in the ZND half reaction length. For the wide duct, the size normalized by the reference length is 16x20x20. To obtain a resolution of 15 points per half reaction length, the total number of grid points is set to 241x301x301. In order to minimize computational expenses, we have employed only every odd grid point in the transverse direction in the calculations, which implies a grid arrangement of 241x151x151. As such, this grid arrangement corresponds to a resolution of 15 points in the ZND half reaction length in the x direction, but up to 8 points in the ZND half reaction length in the *y* and *z* directions. Results obtained employing these two grid arrangements are similar for a given state, and the resolved detonation profiles are close to each other. This is because for the flow in a duct, the main velocity component *u* is in the streamwise direction, the transverse velocity components are small compared to *u* and the gradients of all the flow parameters are also lower in the transverse directions. He and Karagozian [21] carried out simulations of detonation waves using a WENO scheme. Their results suggested that even a grid of 5 points per reaction zone half-length is sufficient for capturing the detonation wave structure. One may note that the use of random initial perturbations generally requires a longer time period before 3D cellular patterns are established. Thus considerable computational time is necessary for the wave to settle to the final quasi-stabilized flow pattern.

### 3.2.1. Narrow duct simulation

Figures 4(a-d) show contour plots of density, pressure, velocity, and reactant mass fraction, respectively, at the initial stage of the disturbance development in the detonation. Initially, the flow is uniform and there is a little change in the transverse directions. The disturbance gradually causes the reacting flow to become unsteady yielding a time dependent flow pattern. The detonation front is correspondingly distorted (Fig.4). After a sufficiently long time, the flow downstream transits to a quasi-steady periodic state in which the distorted front evolves cyclically at the walls. Figure 5 shows the density contour in a period which translates on the four walls in a clockwise direction. The details of the variation of the triple point line can be observed during a cycle. It is found that the period is constant when the flow has reached a quasi-steady state. Figure 6 shows a schematic diagram of the motion of the triple point lines, which is deduced from Fig.5. Figure 7 shows the cell patterns determined by recording the history of the maximum pressure on the side walls. In this figure, the detonation propagates from left to right at an approximately constant sloping angle. An angle with respect to the transverse direction namely the pitch is about 50°. This value is calculated from the streamwise length of the front covered during a period and from the perimeter of the duct. It can be observed from Fig.7 that the streamwise length traveled in a period is about 20*L*. The patterns on the other two opposite walls are similar, and they are therefore not shown here.

The phenomenon of spinning motion about the axial direction in a square tube, similar to the present work has also been recently reported [34]. Other earlier 3D simulations [15-19], made no recollection of such spinning motions. As we shall see in the next section for a larger duct, the spinning motion of the detonation front is not observed. It is sufficient to note that for the selected set of parameters, even in a narrow channel, the detonation at the CJ condition can be obtained. As indicated below for the wider duct, the absence of the spinning motion does not preclude the occurrence of detonation.

Experimentally, the spinning motion of detonations probably first observed in the early 1920's was recently revisited [35, 36], but the mechanisms leading to the spinning motion of a detonation front is not adequately described. Experiments [35] in a circular pipe for a two-phase material system indicate that spinning detonation front exists in the entire cross-section and the angular velocity for the spinning motion is constant. The transverse wave and the "tail" wave become weaker when they approach the center of the cross-section. It was also suggested by the authors that transverse waves probably play a dominant role for stable detonation front propagation. The analytical study for a circular pipe [36] suggests that the spinning detonation originates from a "spinning instability". However, it has been pointed by the authors that the spinning detonation is inherently three-dimensional and hence constitutes a challenging problem. Although the spinning detonations occurring in the circular pipe and rectangular duct are not quite the same, they may share some common features. Thus, the phenomenon of spinning motion of the detonation can only be captured by numerical simulation in 3D dimensions. It can be suggested from these results that there are substantial differences between 2D and 3D detonation calculations.

### 3.2.2. Wide duct simulation

Detonation in a larger width channel is calculated in the same manner as that used for the narrow channel. Again at the start, the flow perturbation is initiated with a random disturbance at the front. It is found that a few small humps are generated at the front with non-uniform amplitudes (Fig.8), and these perturbations become larger with time. It is also found that with the increase of hump size, pressure and velocity behind these humps gradually increase to values much higher than those of ZND values, while in those areas between these humps the pressure and velocity gradually become lower than those corresponding to ZND conditions. As a result, in these latter areas, the detonation becomes weak and the chemical reaction becomes less intense compared to that in the Mach stem. These humps are equivalent to the Mach stems in 2D detonation, while the flat areas among the humps are the incident shock wave plane.

After further evolution, these small humps wander within the domain and then connect together to form larger structures. Simultaneously, transverse waves are formed along the four side walls, as shown in Fig.8. After a considerable time evolution, the detonation front develops with a quasi-steady periodic motion. The detonation front on the four walls moves partly out-of-phase with each other. The detonation front shows a quasi-steady periodical "rectangular mode", with the front displaying a "convex" followed by a "concave" front (Fig.9 and Fig.10). This rectangular front is similar to those reported by Williams et al. [15] and Tsuboi et al. [16]. On the other hand, these simulation results differ from those of Ref. [17], which shows that the phases at neighboring walls are always the same in the "rectangular" mode.

Records of the maximum pressure on two sidewalls (at z=0 and y=0) of the wide duct are displayed in Fig.11. These snapshots pertain to the detonation front in the periodic regime after a very long evolution time. It can be seen that the transverse waves on the neighboring walls are partly out-of-phase and "slapping" waves exist on the two sidewalls. These patterns are similar to those reported by Tsuboi et al. [16] who used a more detailed chemical reaction scheme. It can be observed from Fig.11 that the length of the cell in the streamwise direction is about 40$L$. The present value is the same as those obtained for the same flow and reaction conditions as reported in Ref. 18.

To allow comparison and demonstration of mesh convergence, we simulated a 2D detonation with the same grid resolution and at the same flow and reactant parameters by setting the grid number as one in the $z$ direction. We also recorded the history of maximum velocity and pressure, and compared the cell patterns. It was found that the pattern for the maximum velocity is similar to that of maximum pressure in 3D detonation simulations. This is the case because these patterns represent the trace of the triple points where both the velocity and pressure reach their maxima.

Figures 12 (a) and (b) show distributions of maximum pressure for 2D detonation when the detonation has evolved after a fairly long period and the pattern shows a high degree of quasi-steadiness. The cell width obtained is about 20$L$, which is the same as that found in Ref. 18. The final cell size in the streamwise direction is about 40$L$, which is about the same as that for 3D detonations shown in Fig.11. In general, for 2D detonations, a regular and smaller sized diamond shaped cell pattern is generated in the relatively early stages of its evolution [9-14, 25]. The cell pattern as observed in the early stages then becomes unstable at a later time and finally, the small cells merge into larger cells further downstream [see also Ref. 25].

As mentioned above, earlier experimental works [8] indicate the presence of two types of 3D detonation structures designated as "rectangular" and "diagonal". These experiments also show that there are two types of rectangular structures: in phase and out of phase. The present simulations indicate that the rectangular structure consists of two waves propagating separately in two directions, but the behavior of the detonation in this structure is not a simple overlap of a couple of 2D waves. The resulting detonation structure is essentially 3D, and hence certain aspects such as the cell shape differ from those found in two-dimensional simulations (see Fig.11 versus Fig.12). Typically, the cell pattern from 2D simulation misses the "slapping" waves observed in experiments [8], while the cell pattern from 3D simulation captures this behavior faithfully (Fig.11). As such, again simply investigating 2D detonation may not reveal fully the mechanism occurring in 3D detonations. As a prominent example, the spinning detonation for the narrow duct discussed in Section 4.2.1 will naturally not be captured in 2D simulations (see also ref. 15-19).

**3.3 Detonation physics and role of transverse waves**

The physics of detonation can be considered by comparing results of one, two, and three-dimensional detonations. In 1D detonation (Fig.1), the peak velocity and peak pressure are always well coupled (or closely linked spatially) so detonation can be readily sustained. This is also true for the case in 3D detonation at the initial stages when disturbances are introduced but have not spread yet. In this situation, the detonation front is essentially a plane perpendicular to the streamwise direction. The peak pressure attained is the ZND pressure and the peak velocity takes the CJ velocity, as shown in Figs. 1 and 2. All of the fuel after the detonation front layer has reacted, and there is no fresh fuel pocket formed. Uniform detonation occurs along the entire front. When the peak pressure becomes less than the ZND pressure, the maximum pressure and the maximum velocity are not well coupled. This will eventually lead to "decoupling" of the shock wave and the chemical reaction regions. The peak pressure and peak velocity are produced by the supersonic shock wave formed and the energy release produced by the reacted fuel. Should there be no addition of energy by the combusted reactants, the detonation front degrades to a pure shock wave.

In 2D and 3D detonations, transverse waves are formed between the regions of differing reaction rate near the front. As such, the distribution of detonation strength is not uniform along the front with the detonation being stronger around areas where high pressure and high velocity occurs behind the convex Mach stem and at the triple points. Thus, "hot spots" are generated in the Mach stems cores and at the triple points. At the triple point, the maximum pressure and maximum velocity are well coupled like in 1D detonation, but their values can be larger than those arising in 1D detonation.

However, in some areas along the front (incident shock), the detonation is weak. It is found that peak pressure and peak velocity values along the *incident shock* are much lower than those of the ZND values and the associated reaction in such locations is less intense than that near the triple points. As the incident shock moves into the un-reacted gas region, the fuel passing this incident shock might not be completely combusted. That is the reason why there are some un-reacted fuel pockets due to the uneven detonation front, which has also been observed in experiments [37]. Along the incident shock on the front (which is the area between Mach stems), if there is no further heat production, the detonation will decay. It turns out that these areas are eventually swept by the motion of the triple points with an associated increase of temperature, which induces high reaction rates that causes rises in pressure and velocity. Thus, it is suggested that the detonation in 2D and 3D can be described as being sustained by the transverse motion of hot spots along the detonation front. Detonation occurrence depends on the close coupling of the maximum pressure and velocity.

On the role of transverse waves, some researchers hold the view that these waves play an important part. It is believed that transverse waves are necessary if a steady detonation is to be sustained [13, 4]. However, a question arises as to how detonation can be sustained in 1D where clearly there is no transverse wave present as shown in Fig.1. This is also the situation as observed at the beginning of the computation in our 3D simulations before the disturbance leads to transversal variations. Thus, it would seem that transverse waves per se are not the inherent mechanism for detonation. However, undoubtedly the motion of transverse waves leads to triple point tracks that produce regions of high pressure and large velocity (hot spots), and hence enables a close coupling of the maximum pressure and velocity. This strong coupling in the area of triple points makes the detonation locally overdriven which further maintains the detonation. Thus, it can be concluded that detonation in 2 and 3 dimensions is sustained by the transverse motion of hot spots over the detonation front.

In the design of pulse detonation engines, the mixture parameters impose limits for detonation to occur [38]. The cell size is one of the critical factors to consider when selecting the tube diameter in such systems. It is known that detonation typically cannot propagate in a tube of diameter less than a critical size; there seems to be a minimum diameter which allows sustained propagation [38]. For a given pipe's diameter, the detonation will be stronger if the channel contains more cells. In other words, the detonation is easier to sustain when the cell size is smaller for a given pipe.

This effect of cell size on the detonation can be explained from our above observations. When the cell size is smaller, there are (possibly) more triple points per unit length along the transverse direction, thus ensuring the presence of high pressure regions along the front. Hence, a strong detonation can be generated which in turn reinforces the presence of more cells in the channel. Under the same conditions, a larger number of triple points per unit length implies that the length of incident shock along the front becomes shorter. A given location at the front will then be swept by hot spots at a higher frequency. This causes the pressure and velocity along the front tend to be high. This in turn increases the reaction strength at the front, which avoids or minimizes the presence of un-reacted fuel pockets.

In addition, Hu et al. [39] provide an analytical study of 2D detonation front, which showed that the detonation wave is, first, strengthened at the front of the cell after the triple-shock collision, and then decays when reaching the cell end. Their analysis seems to agree with the simulation results and experimental observations. To further assess the detonation mechanism, our 2D simulation results are shown in Fig.13. Here this figure depicts the smaller size cell pattern at the early stages of the detonation evolution. In this case the contour of maximum pressure exceeding the ZND pressure and the contour of maximum velocity larger than the CJ velocity are also shown in Fig.13 (c) and (d), respectively. It can be seen that the pressure in the second half cell is generally less than the ZND pressure (after being passed by the incident shock wave), and the velocity in the second half cell is also less than its CJ value. As a result, in these areas, detonation becomes weaker. Reaction in the second half cell is heated by the triple points via the transverse motion on the front. This observation is in agreement with Hu et al's analysis [39]. If the overdrive factor is near unity, the detonation may also be easily quenched or degraded particularly for cases where the cell sizes per unit length are larger. Pintgen et al. [20] noted that the detonation is always more unstable when the overdrive factor is near unity compared to cases of higher overdrive factor. When the overdrive factor is near unity, the detonation is not strong at the front, and the cell pattern is only marginally sustained, in particular, for large cell sizes in a given duct.

Finally, to provide further observations to support the suggestion that the detonation mechanism is attributed to the coupling of maximum pressure and maximum velocity, one can compare the contours of maximum pressure and maximum velocity between the Mach stem area and the incident shock area along the front as displayed in Fig.13. In the core area of the second half of the cell, although the maximum velocity is still larger than the CJ value, the maximum pressure has become much lower than the ZND value. In general, when the pressure is low, the chemical reaction is retarded. As a result of this decoupling of the velocity (greater than CJ value) with the pressure (greater than ZND value), reaction in the second half of the cell is less intense. Also from Fig.14, one can distinguish the features between the Mach stem and the incident shock by the relative thickness of the reaction zone. Along the Mach stem, with the close coupling of maximum velocity and maximum pressure (as found in the first half cell in Fig.13), the reaction zone is thin, and the gradients of flow and reaction variables are comparatively steep. The reaction is completed in a shorter distance (see Fig.14). On the other hand, along the incident shock the reaction zone is relatively thicker, and the reaction takes a longer time to complete. As a result, the centre where the reaction occurs moves a little further from the leading shock front as compared to Mach stem area. This can lead to the decoupling of the reaction zone and the leading shock. This is particularly prominent if the cell size is large.

## 4. Conclusions

Numerical simulations of propagating detonation waves in 3D geometries have been presented, and the chemical reaction is represented by a single step Arrhenius law. The Eulerian conservation equations are integrated using a 3D TVD Runge-Kutta time stepping method with a fifth-order WENO scheme for spatial discretization. To better examine the actual physical process, only via 3D calculations can one hope to obtain the full features of detonation flow structure. Results indicate that there are important differences between three-dimensional and two-dimensional detonations, though they also share common features.

The three-dimensional effect of the detonation wave front comes to the fore when one considers detonation in ducts of different dimensions. Calculations in a narrow square duct using a grid of sufficient resolution (where its transverse size is a fraction of the cell width) clearly shows a sustained detonation wave front which spins about the axial direction of the duct. The transverse wave formation switches between the four sidewalls in a cyclic manner.

For the larger duct, the quasi-steady motion of the front is of a different type and the spinning motion is not present. The front shows a quasi-steady periodic "rectangular mode," with alternate "convex" and "concave" front shapes. For the same duct width, flow and reactant parameters, the pattern of maximum pressure on one of the side walls of the 3D detonation differs from that obtained from the 2D simulations. But, the length of the cell in the streamwise direction as deduced from the pressure pattern in the 3D simulation is about the same as that for the 2D simulation.

The present work indicates that the reaction process is dominated by "hot spots" regions that features high pressure and high velocity, i.e., triple points and the core of Mach stem sweeping over the wave front in the transverse direction for both the 2D and 3D detonation simulations. The motion of transverse waves leads to tracks of triple points which induce zones of high pressure and large velocity coupled together. In these regions the detonation is locally overdriven which in turn maintains the detonation along the wave front. However, transverse waves are not a necessity for sustaining a detonation as demonstrated by the 1D detonation simulations, but it is important to have strongly coupled high pressure and high velocity regions. The present results show that in the 2D and 3D detonation simulations, owing to the disturbance and the unsteady behavior of reacting flows, transverse waves indeed occur. These transverse waves facilitate the coupling of high pressure and high velocity regions for enhanced chemical reaction.

**Acknowledgements**

The authors would like to thank Profs. Z. Jiang and L. Bauwens, and Drs. X. Hu and C. J. Teo as well as Mr. C. Turangan for their helpful discussions. The authors would like to acknowledge the research funding provided by the Singapore Defense Science and Technology Agency for the study. Useful discussions with Dr. Y. T. Jiang of DSO National Laboratories and their support for the project are also appreciated.

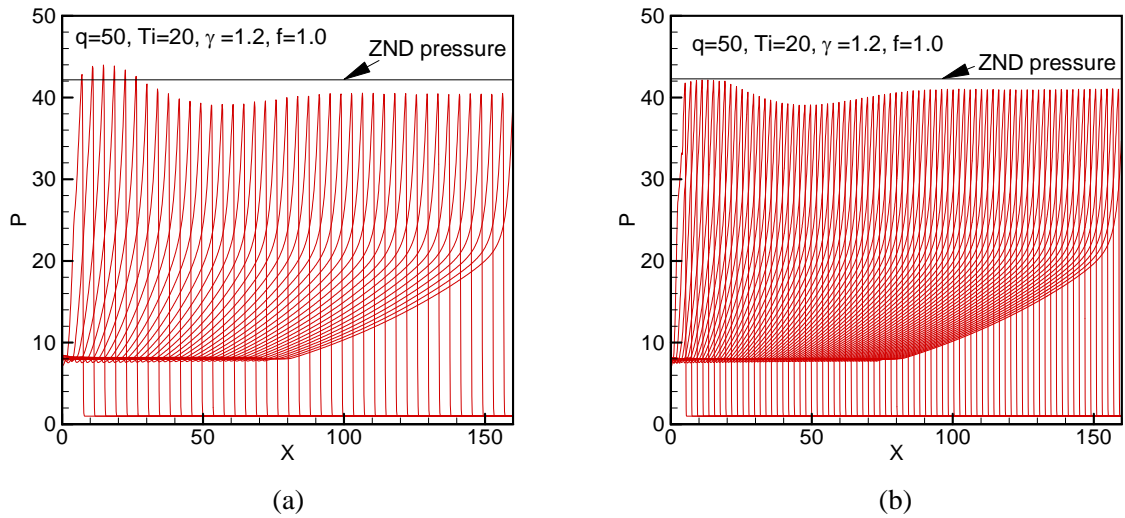

Fig.1 Pressure profiles in simulations of one-dimensional detonation wave propagation. (a) Resolution is 8 with points per half-reaction length. (b) Resolution is with 16 points per half-reaction length.

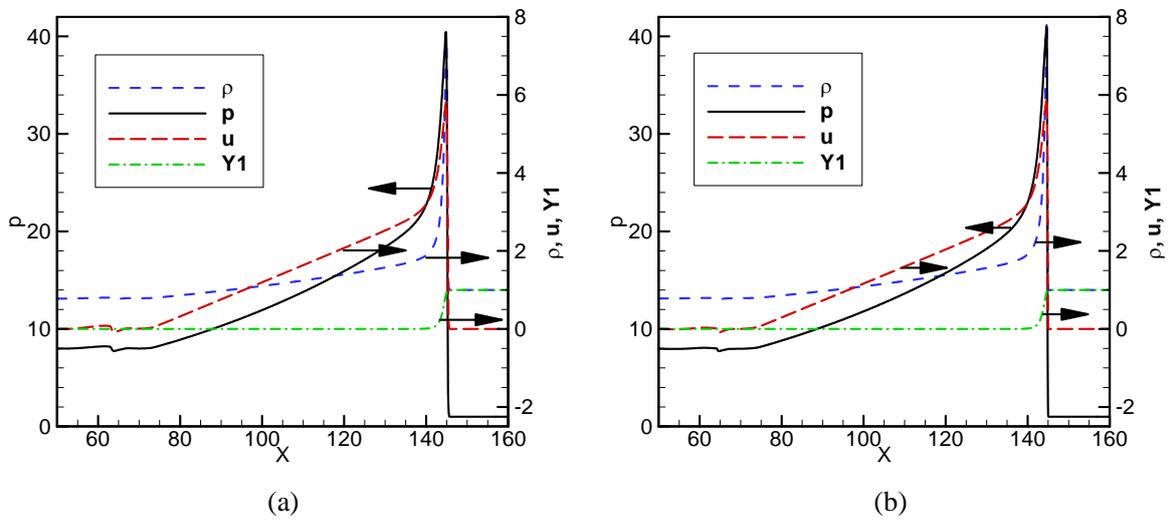

Fig.2 Flow variable profiles of steady solution of one-dimensional detonation for Q=50, Ti=20, γ=1.2, and f=1.0. Here, Y1 refers to the reactant fraction Y. (a) Resolution is with 8 points per half-reaction length. (b) Resolution is with 16 points per half-reaction length.

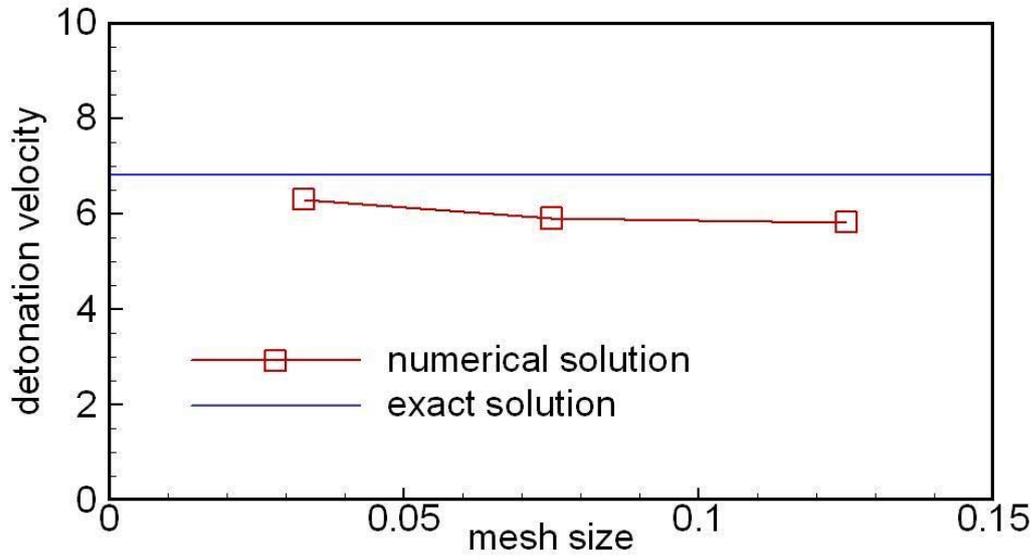

(a)

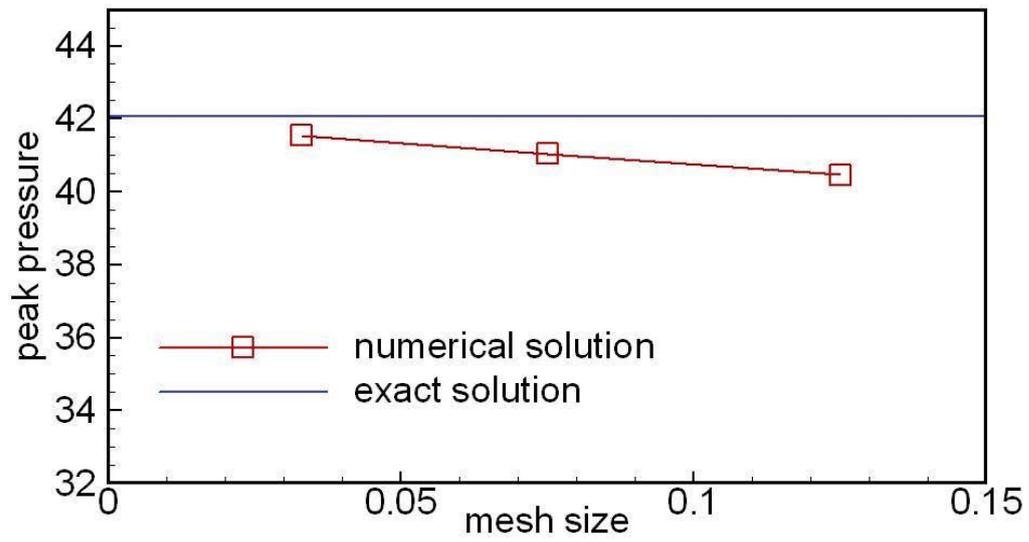

(b)

Fig.3 Mesh convergence for the simulation of one-dimensional detonation. (a) detonation velocity; (b) peak pressure.

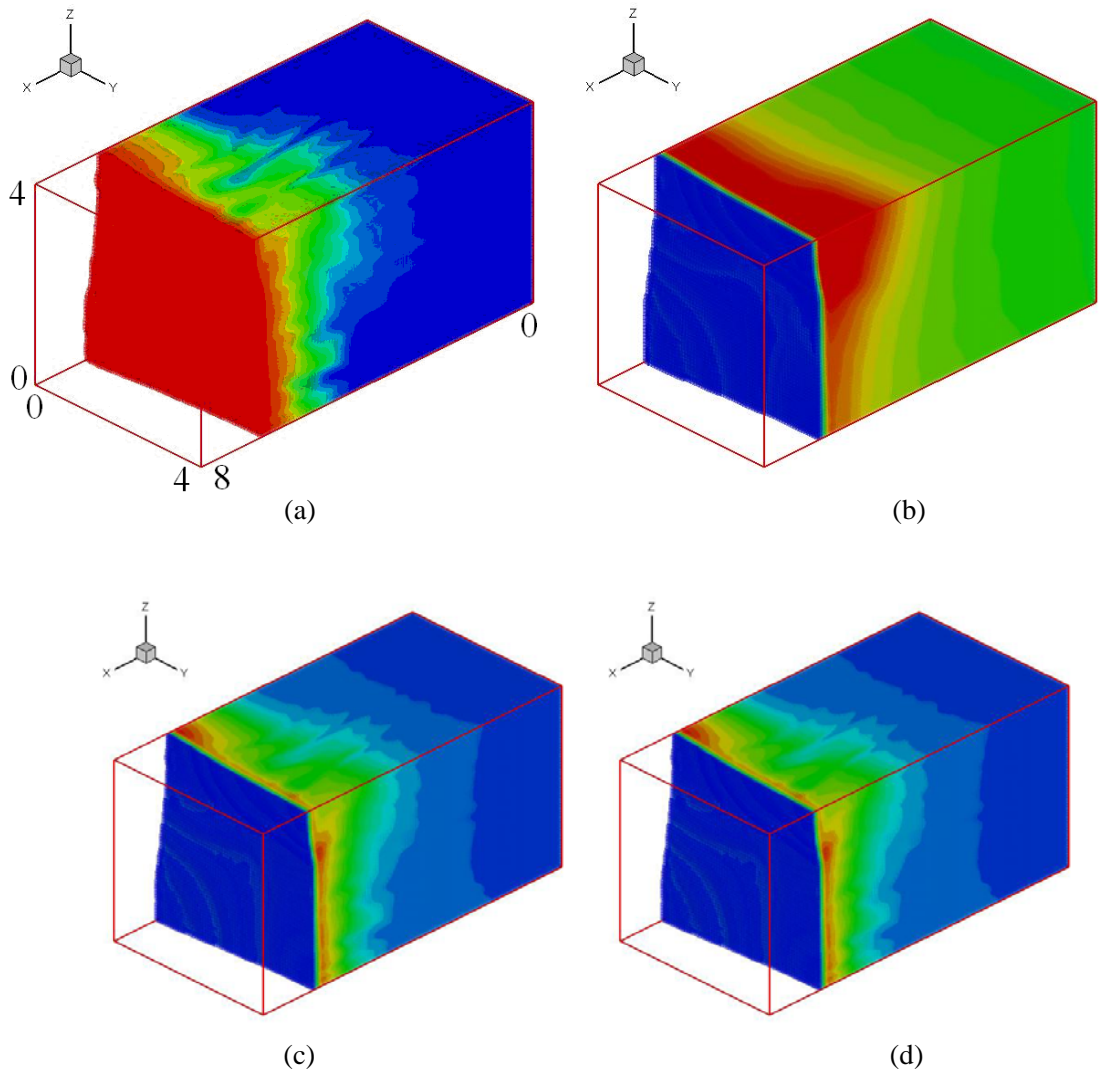

Fig. 4 Contours of the flow and reaction variables at 60,000 time steps for the narrow duct at dimensionless time $t$=22.26. The parameters for the reaction are $q$=50, $T_i$=20, $\gamma$=1.2, and $f$=1.0. (a) mass fraction of reactant; (b) pressure; (c) density; (d) streamwise velocity.

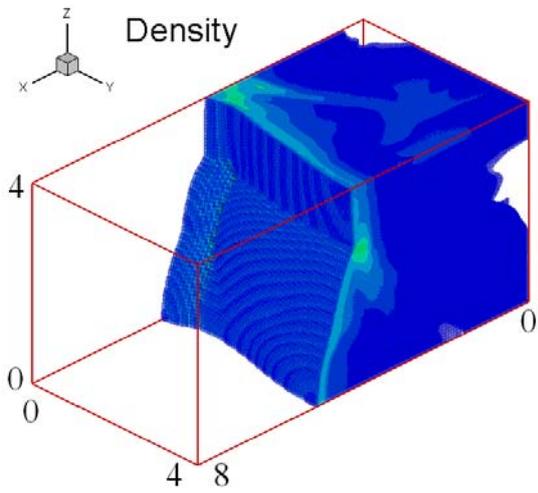
(a)

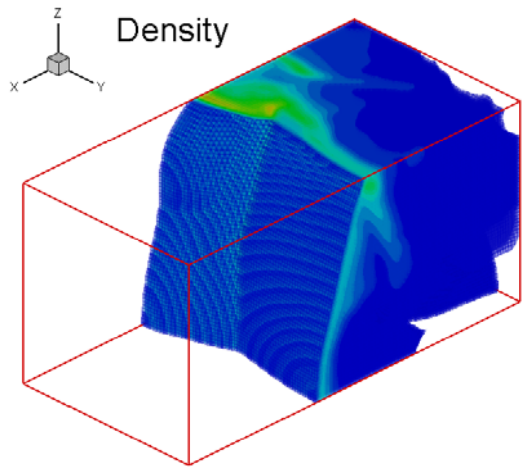
(b)

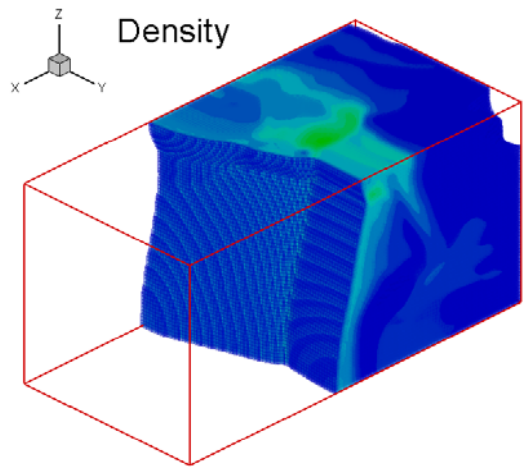
(c)

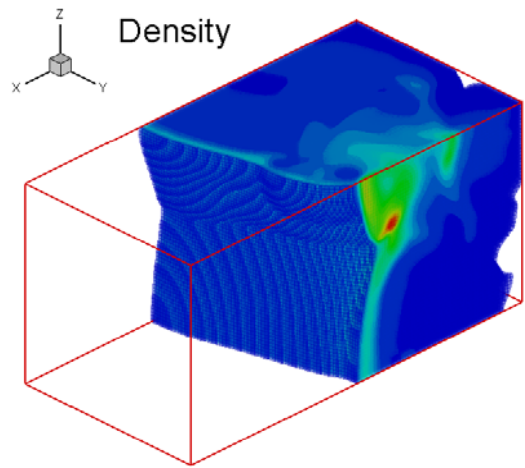
(d)

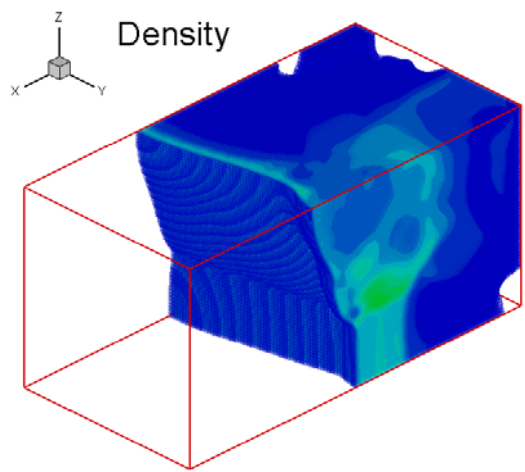
(e)

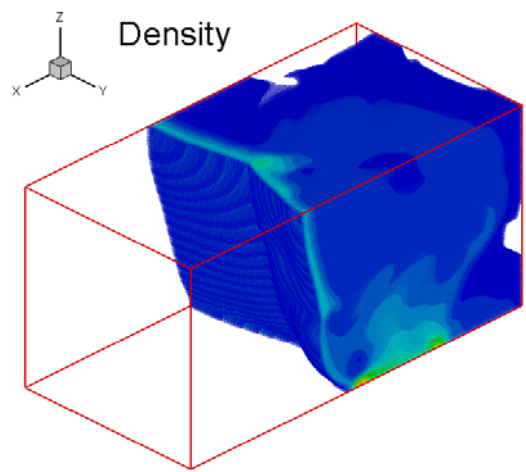
(f)
23

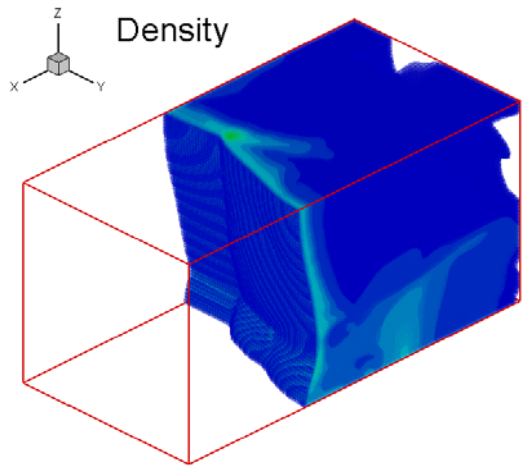 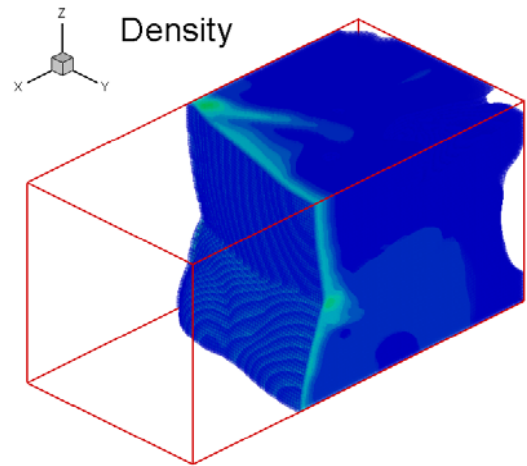

(g) (h)

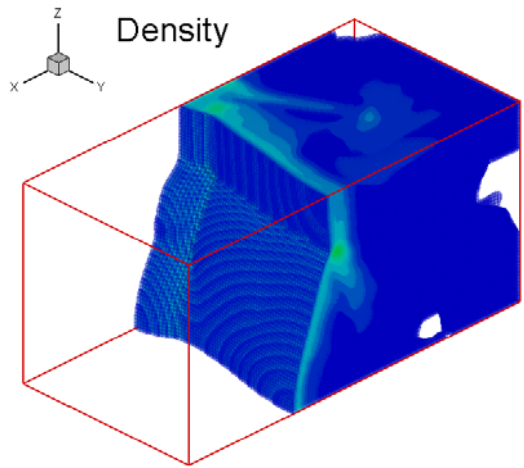

(i)

Fig.5 Evolution of the detonation front on the walls of the narrow duct for a full period. The flame front spins/moves in a clockwise manner. Frames (a) – (i) are equally spaced with the same number of time steps.

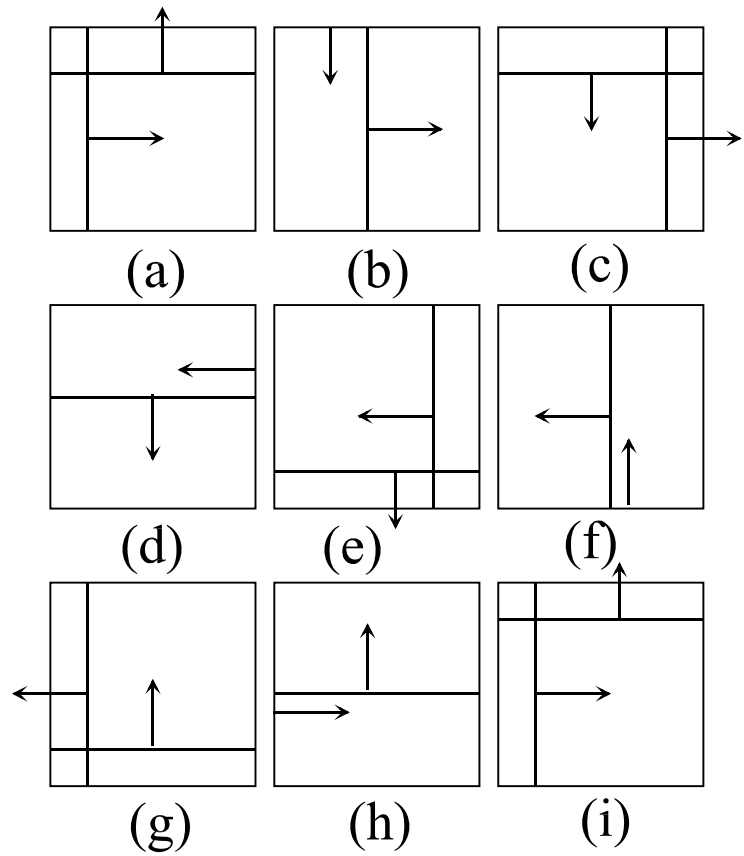

Fig.6 Schematic of movement of the triple point line in a period of spinning.

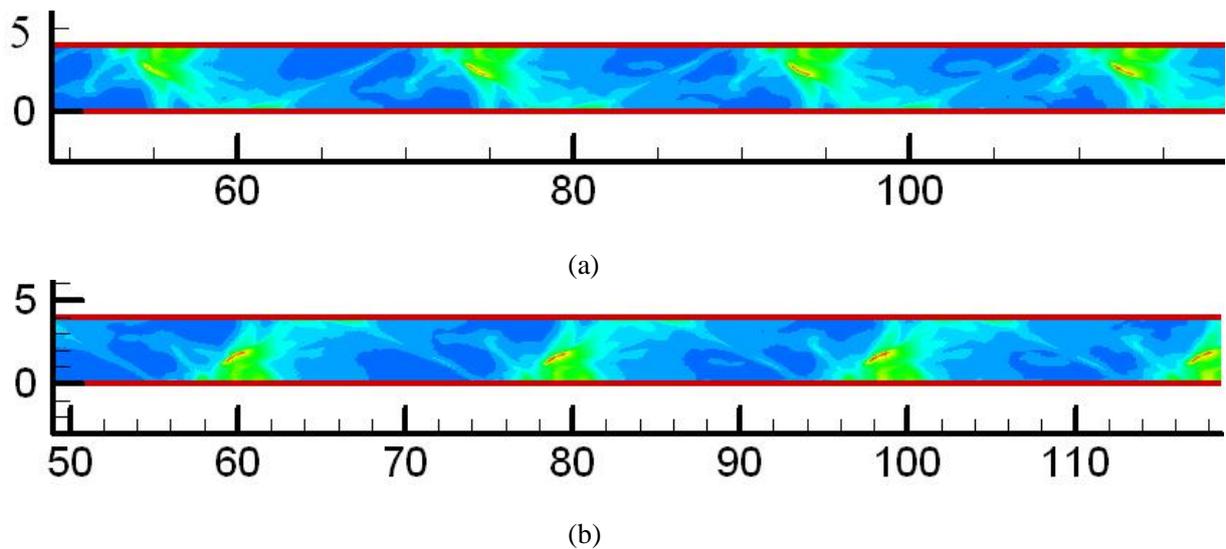

(a)

(b)

Fig. 7 History record of the maximum pressure on the walls (detonation propagates from left to right). (a) Side wall at z=0; (b) Side wall at y=0. The spinning angle measured from the transverse direction is 50.1 degree.

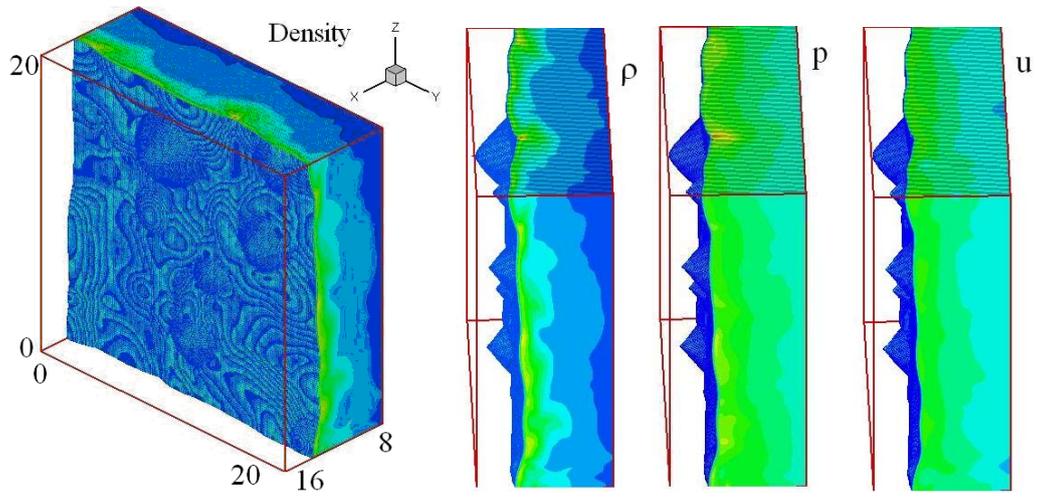

Fig.8 Formation of humps at the detonation front in the wide duct at the starting stage of 3D simulation.

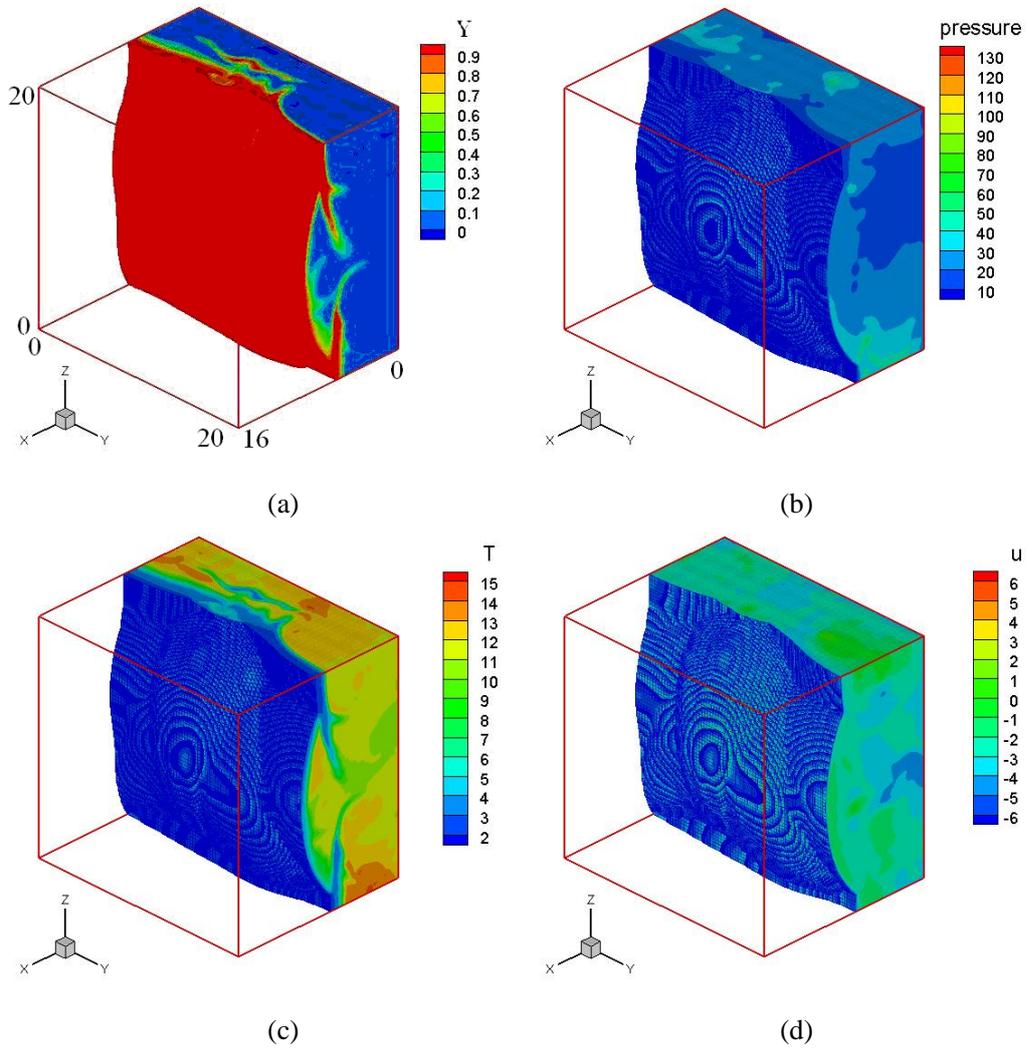

Fig. 9 Contours of various parameters for the wide duct at $t=46/140$ in a period of time 140. (a) mass fraction of reactant; (b) pressure; (c) temperature; (d) streamwise velocity.

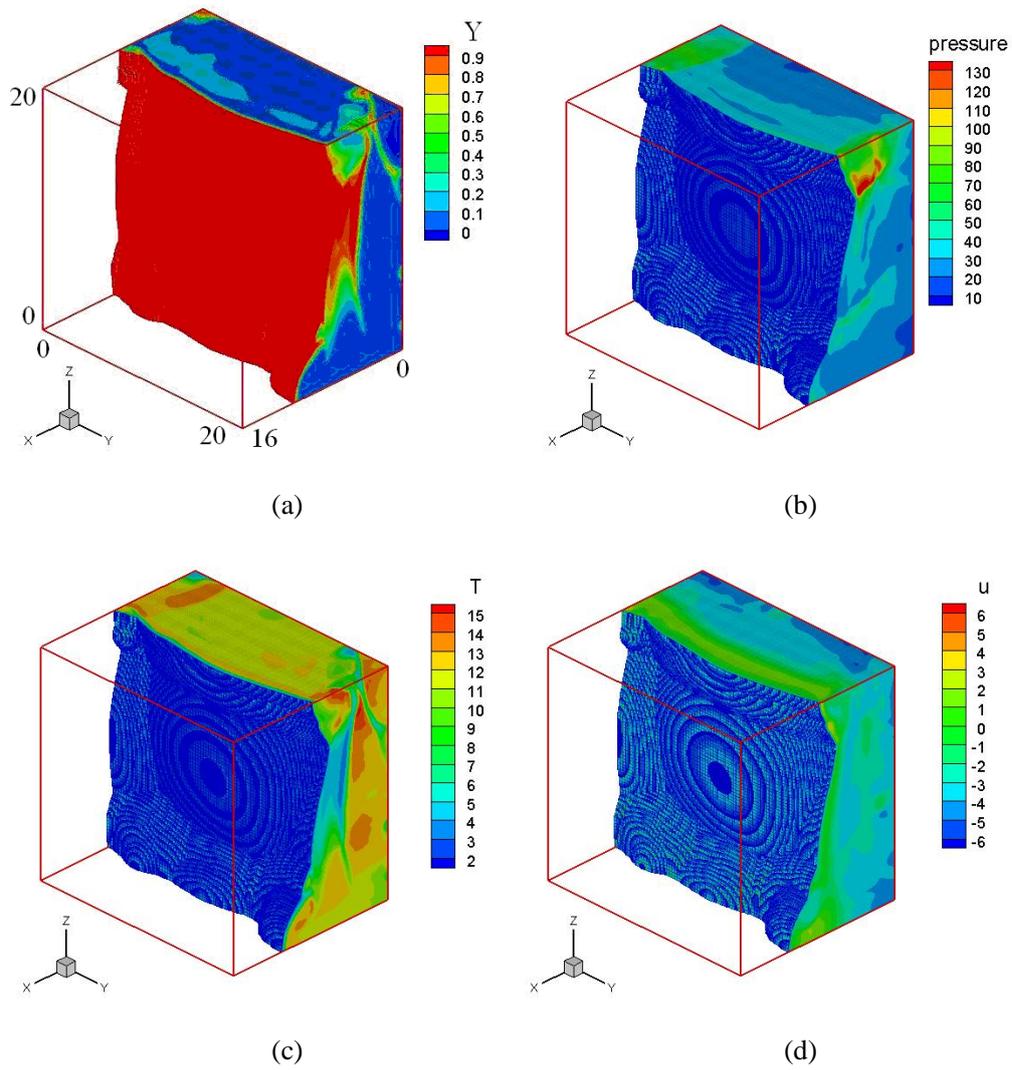

Fig.10 Contours of various parameters for the wide duct at t=94/140 in a period of time 140. (a) mass fraction of reactant; (b) pressure; (c) temperature; (d) streamwise velocity.

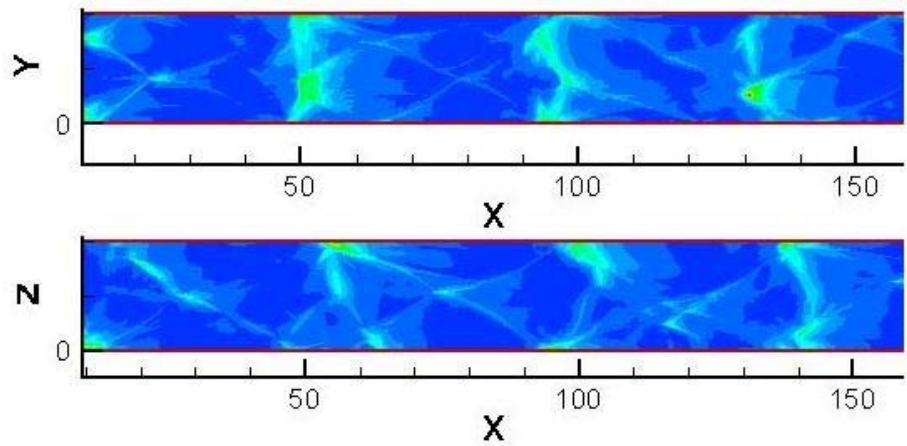

Fig. 11 History of maximum pressure on side walls of the wide duct of 3D detonation after the detonation has settled into the quasi-steady state. Top: On wall z=0; Bottom: On wall y=0.

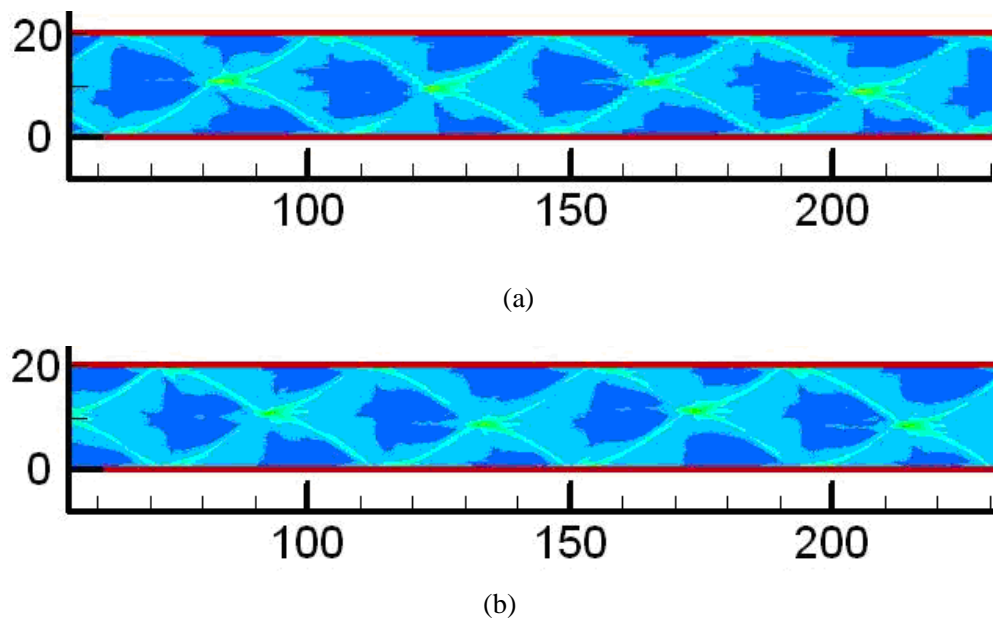

(a)

(b)

Fig. 12 History of maximum pressure for 2D detonation in the wide duct for Q=50, Ti=20, $\gamma$=1.2, and f=1.0. (a) Resolution is with 16 points per half-reaction length. (b) Resolution is with 32 points per half-reaction length.

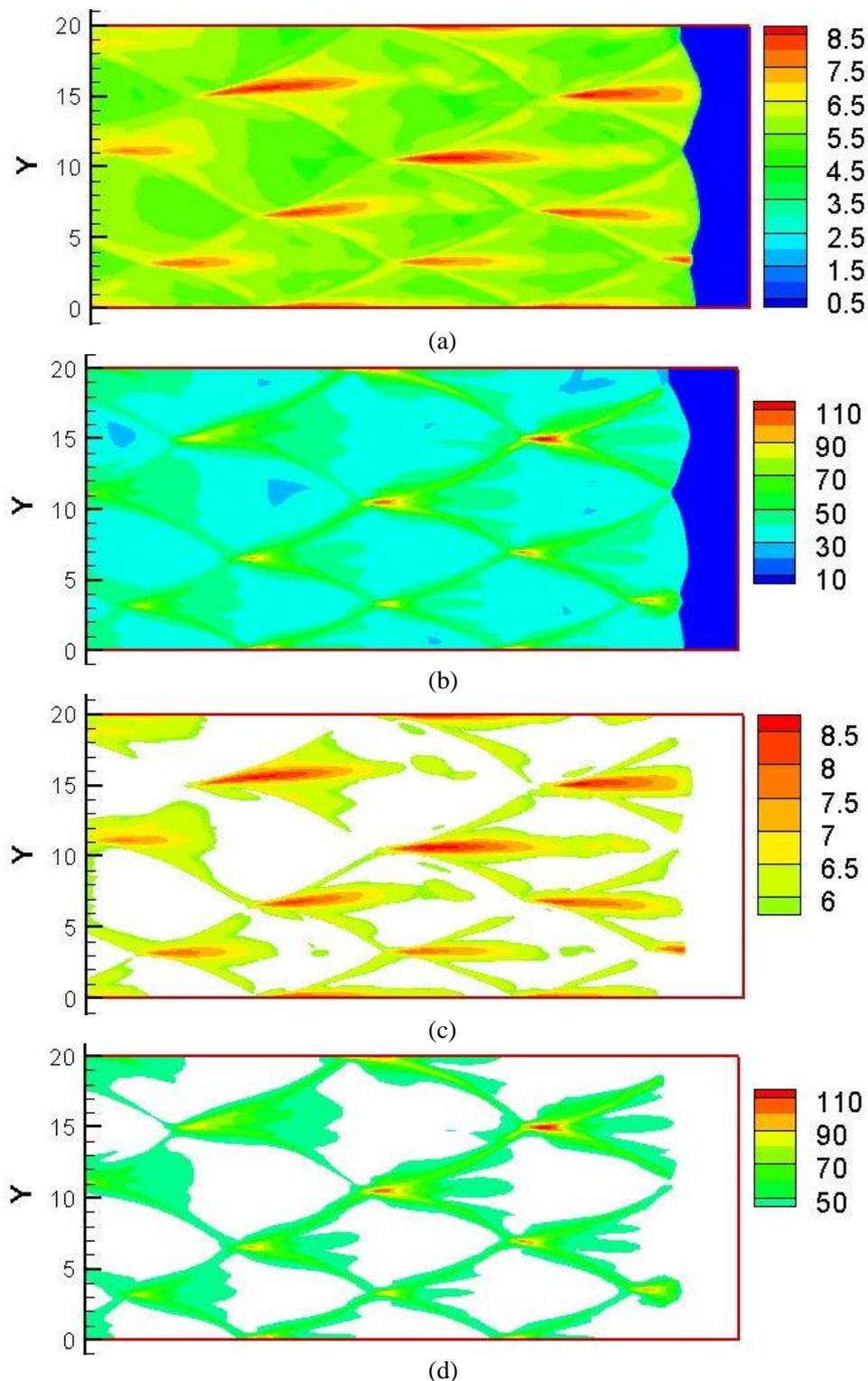

Fig.13 Cell pattern formed in the early stage for the two-dimensional detonation simulations for the wide duct size. (a) History of the maximum velocity contour; (b) History of the maximum pressure contour; (c) History of the maximum velocity contour depicting areas of velocity larger than the CJ value. (d) History of the maximum pressure contour depicting areas of pressure larger than the ZND value.

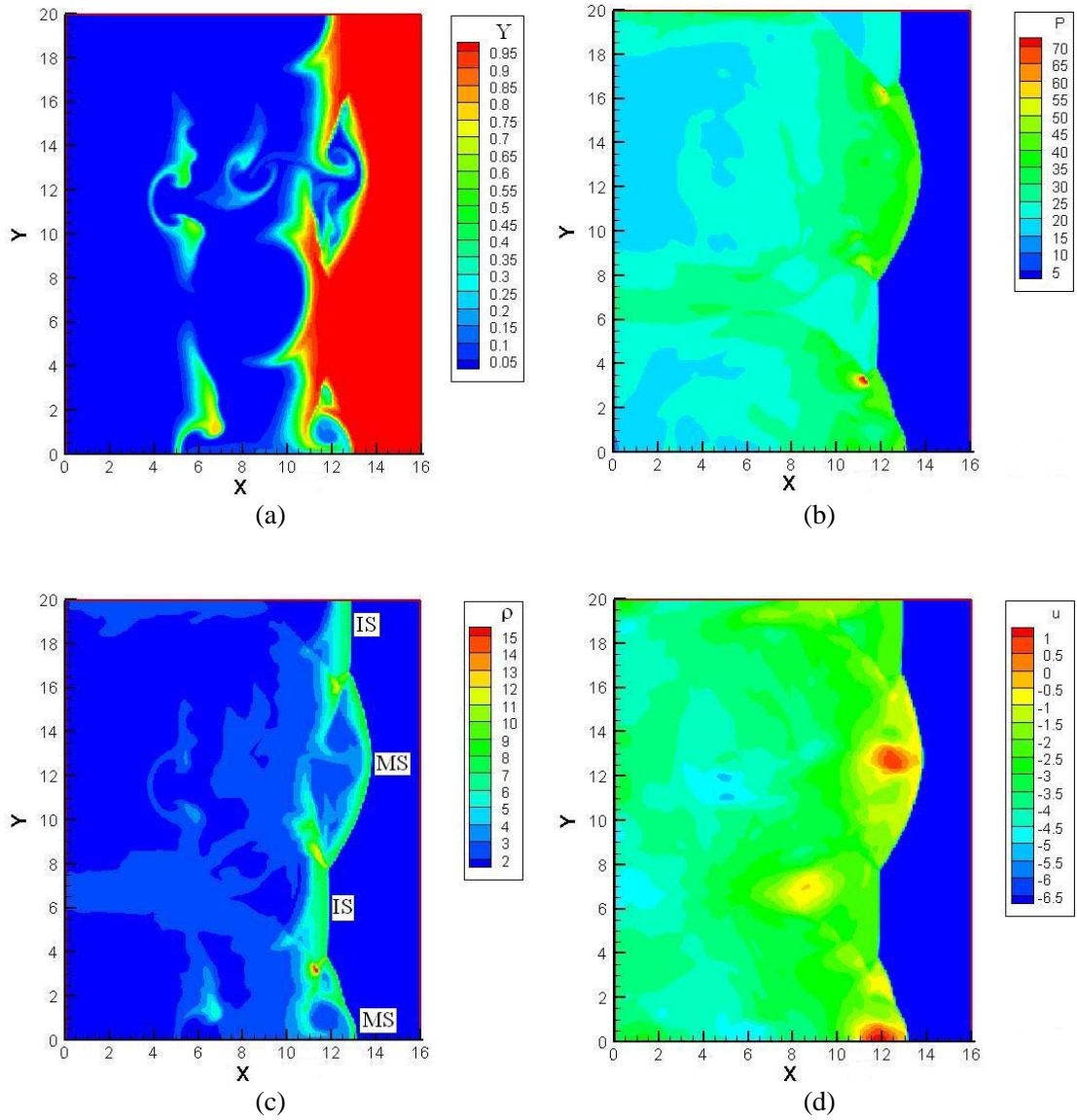

Fig. 14 Contours of the flow and reaction variables in the early stage for the two-dimensional detonation simulations for the wide duct size. The parameters for the reaction are q=50, $T_i$=20, $\gamma$=1.2, and f=1.0. (a) mass fraction of reactant; (b) pressure; (c) density; (d) streamwise velocity. In the picture of density, MS and IS stand for Mach stem and incident shock, respectively.